\documentclass[preprint2]{aastex}
\usepackage{natbib}

\usepackage{color}

\begin{document}

\title{Quantifying the Impact of Spectral Coverage on the Retrieval of Molecular Abundances from Exoplanet Transmission Spectra}

\author{John W. Chapman\altaffilmark{1}, Robert T. Zellem\altaffilmark{1}, Michael R. Line\altaffilmark{2}, Gautam Vasisht\altaffilmark{1},  Geoff Bryden\altaffilmark{1}, Karen Willacy\altaffilmark{1}, Aishwarya R. Iyer\altaffilmark{1,3},
Jacob Bean\altaffilmark{4},
Nicolas B. Cowan\altaffilmark{5},
Jonathan J. Fortney\altaffilmark{6},
Caitlin A. Griffith\altaffilmark{7},
Tiffany Kataria\altaffilmark{1},
Eliza M.-R. Kempton\altaffilmark{8},
Laura Kreidberg\altaffilmark{9},
Julianne I. Moses\altaffilmark{10},
Kevin B. Stevenson\altaffilmark{11},
Mark R. Swain\altaffilmark{1}}

\affil{$^{1}$ Jet Propulsion Laboratory, California Institute of Technology, 4800 Oak Grove Dr, Pasadena, CA 91109, USA}
\affil{$^{2}$ School of Earth \& Space Exploration, Arizona State University, USA}
\affil{$^{3}$ California State University, Northridge, 18111 Nordhoff Street, Northridge CA 91330, USA}
\affil{$^{4}$ Department of Astronomy \& Astrophysics, University of Chicago, 5640 S Ellis Avenue, Chicago, IL 60637, USA}
\affil{$^{5}$ Department of Earth \& Planetary Sciences, McGill University, 3450 rue University, Montréal, QC H3A 0E8, Canada}
\affil{$^{6}$ Department of Astronomy \& Astrophysics, University of California, Santa Cruz, CA, USA}
\affil{$^{7}$ Lunar and Planetary Laboratory, University of Arizona, 1629 East University Boulevard, Tucson, AZ 85721, USA}
\affil{$^{8}$ Department of Physics, Grinnell College, 1116 8th Ave., Grinnell, IA 50112, USA}
\affil{$^{9}$ Harvard-Smithsonian Center for Astrophysics, 60 Garden Street, Cambridge, MA 02138, USA}
\affil{$^{10}$ Space Science Institute, 4750 Walnut Street, Suite 205, Boulder, CO 80301, USA}
\affil{$^{11}$ Space Telescope Science Institute, 3700 San Martin Drive, Baltimore, MD 21218, USA}

\begin{abstract}
Using forward models for representative exoplanet atmospheres and a radiometric instrument model, we have generated synthetic observational data to explore how well the major C- and O-bearing chemical species (CO, CO$_2$, CH$_4$, and H$_2$O), important for determining atmospheric opacity and radiation balance, can be constrained by transit measurements as a function of spectral wavelength coverage. This work features simulations for a notional transit spectroscopy mission and compares two cases for instrument spectral coverage (wavelength coverage from  0.5$-$2.5~$\mu$m and 0.5$-$5~$\mu$m).  The simulation is conducted on a grid with a range of stellar magnitudes and incorporates a full retrieval of atmospheric model parameters. We consider a range of planets from sub-Neptunes to hot Jupiters and include both low and high mean molecular weight atmospheres. We find that including the 2.5--5~$\mu$m wavelength range provides a significant improvement in the degree of constraint on the retrieved molecular abundances: up to $\sim$3 orders of magnitude for a low mean molecular weight atmosphere ($\mu=2.3$) and up to a factor of $\sim$6 for a high mean molecular weight atmosphere ($\mu=28$). These decreased uncertainties imply that broad spectral coverage between the visible and the mid-infrared is an important tool for understanding the chemistry and composition of exoplanet atmospheres. This analysis suggests that the JWST/NIRSpec 0.6$-$5~$\mu$m prism spectroscopy mode, or similar wavelength coverage with possible future missions, will be an important resource for exoplanet atmospheric characterization.
\end{abstract}



\section{Introduction}
While both ground- and space-based telescopes have probed transiting exoplanets' atmospheric thermal and molecular structures, dynamics, and scattering properties, space-based telescopes do not suffer from limitations imposed by Earth's atmosphere, such as telluric absorption in the infrared. Infrared wavelengths ($0.7-5~\mu$m) are crucial for characterizing the atmopsheric compositions of exoplanets as they probe strong rotational-vibrational  molecular bands of H$_{2}$O, CH$_{4}$, CO, and CO$_{2}$ \citep{seager00, brown01, charbonneau02, Vidal-Madjar03, knutson07,tinetti07, swain08, linsky10, snellen10, majeau12}. However, current infrared transiting exoplanet space-based telescopes are limited to spectroscopy between $0.7$ and $1.7~\mu$m with \emph{Hubble}/STIS and WFC3 \citep{riley17, dressel17} and photometry at 3.6 and 4.5~$\mu$m with \emph{Spitzer}/IRAC \citep{werner04, fazio04} and hence lack the spectroscopic capabilities to measure the strongest and most easily identifiable absorption bands of CO, CO$_{2}$, and CH$_{4}$ \citep{fortney05}. These molecules, along with H$_2$O, are the dominant C- and O-bearing species in exoplanetary and planetary atmospheres and are therefore necessary to estimate robustly an exoplanet's carbon-to-oxygen (C/O) ratio and metallicity \citep{Moses13, venot15, espinoza16}, which provide important constraints on formation mechanism \citep{lodders04, mousis09, oberg11, alidib14, Helled14, Madhusudhan14, Marboeuf14, Madhusudhan16, Mordasini16}. 

Determining the abundances of the O- and C-bearing molecules in exoplanetary atmospheres is of paramount importance as neither the accreted nebular gas, nor the nebular solids, are expected to share the same abundances of the parent star. Snow lines, areas where H$_2$O, CO$_2$, and CO are capable of condensing within the solar nebula, may dramatically alter the C/O ratio of the gas and solids, leading to the dichotomy between molecular abundances of the host star and nebular debris \citep[e.g.,][]{Madhusudhan14, alidib14,Thiabaud15,oberg11, oberg16}. For instance, the condensation of H$_{2}$O at the H$_{2}$O snow line depletes oxygen from the gas phase and enhances oxygen in the solids \citep{oberg11}. Additional condensation of, for example, CO and CO$_{2}$ beyond the H$_{2}$O snow line further partitions the oxygen and carbon in the solid and gas forms \citep{oberg11}. However as a result of, for example, planet migration, connections between atmospheric abundances and accretion disk compositions involve an appreciable level of uncertainty \citep{Mordasini16}. Measurements of the C- and O-bearing molecules in exoplanetary atmospheres therefore provide a method for probing the formation and evolution of planetary systems. However, due to the limitations of current instruments, only a handful of hot Jupiter planets have informed the catalog of exoplanet C/O ratios thus far \citep{Brogi14,kreidberg14a,Brogi15,kreidberg15,line16}.



Here we explore the degree to which an exoplanet's atmospheric molecular abundances can be retrieved as a function of the spectral coverage, stellar magnitude, and planet type.  In contrast, previous studies have assessed the information content of the various instrument modes of NASA's James Webb Space Telescope (JWST) to characterize the atmospheres of transiting hot Jupiters \citep{Batalha17,Howe17}. Here we expand upon them by examining the effect of wavelength coverage on constraining the atmospheric abundances of a H/He-dominated hot Jupiter, a H/He-dominated warm Neptune, a N$_{2}$-dominated warm Neptune, a H/He-dominated sub-Neptune, and a N$_{2}$-dominated sub-Neptune, in a way that is agnostic of the telescope platform. Therefore our study can help guide future observations, such as with JWST which will have the capability to measure transiting exoplanets from 0.6--28~$\micron$ \citep{cowan15,stevenson16}, and mission currently in development, such as ARIEL \citep{puig16,tinetti16} and FINESSE \citep{bean17}.

\section{\label{sec:inst}Telescope and Instrument Model}

Hubble/WFC3 has revolutionized transit spectroscopy by demonstrating the tremendous scope for scientifically useful measurements with relatively low spectral resolution (typically binned to R$\approx$60 or lower at 1.4~$\micron$) and limited wavelength coverage \citep[e.g.,][]{sing16,iyer16,stevensonhst16}. We also consider here the performance of a spectrograph with
modest resolution (R$>$34 at 1.4~$\micron$) but with continuous visible (0.5~$\micron$) to short-infrared (2.5~$\micron$) or mid-infrared (5~$\micron$) wavelength coverage. These combinations of instrument capabilities represents a balance between spectral resolution, sensitivity, and wavelength coverage as well as approximating JWST/NIRSpec's wavelength coverage \citep[0.6--5~$\micron$;][]{rauscher14}.






\begin{deluxetable}{lccc}
\tablewidth{0pt}
\tablecolumns{4}
\tablecaption{Parameters for the two radiometric models. \label{tab:rm}}
\tablehead{
\colhead{Parameter} & \colhead{Model A }&  & \colhead{Model B} 
}
\startdata
\textbf{Spectograph Parameters} & & & \\
Detector wavelength range & $0.5-2.55~\micron$ &$\neq$ & $0.5-5.0~\micron$ \\
Spectrometer temperature & 155 K & $\neq$ & 85 K \\
Integration time & 3600~s &  & 3600~s\\
Detector temperature & 95~K & & 70~K  \\
Detector transmission & 1 &$\neq$ & 0.95 \\
Pixel size & 18~$\mu$m & & 18~$\mu$m \\ 
Detector dark current & 0.5 e$^{-}$/s/pixel & & 0.5 e$^{-}$/s/pixel \\
Detector read noise & 16 e$^{-}$/double read & & 16 e$^{-}$/double read \\
Read interval & 30~s & & 30~s \\
 & & & \\
\textbf{Initial Telescope Parameters} & & &\\
Telescope diameter  & 0.4~m & & 0.4~m \\
Telescope temperature & 220 K &$\neq$ & 119 K  \\
f/\# & 14.4 & $\neq$ & 11.6  \\
 & & & \\
\textbf{Host Star Parameters} & & & \\
Stellar type & K5 &  & K5 \\
Stellar temperature &  4480~K & & 4480~K\\
\enddata
\end{deluxetable} 

\begin{deluxetable}{lcc}
\tablewidth{0pt}
\tablecolumns{3}
\tablecaption{\label{tbl-1}Types of planets simulated}
\tablehead{
\colhead{Planet} & \colhead{Mean Molecular} & \colhead{Terminator} \\
\colhead{Type} & \colhead{Mass $\mu$ (amu)} & \colhead{Temperature (K)} 
} 
\startdata
Hot Jupiter & 2.3 &  1500\\
Warm Neptune &  2.3 & 700\\
Warm Neptune & 28 &  700\\
Sub-Neptune &  2.3 & 600\\
Sub-Neptune & 28 &  600\\
\enddata
\end{deluxetable}

\subsection{Spectrograph}
The models represent the performances of two notional, purpose-built spectrographs (Table~\ref{tab:rm}) featuring the use of a prism as a dispersing element, which provides higher stability, broader wavelength coverage, and total optical throughput ($\sim$0.75, based on design activity for a potential mission) compared to a grism spectrograph  \citep[e.g., Hubble/WFC3's G102 and G141 grisms have a maximum throughput of $\sim$0.4 and $\sim$0.5, respectively;][]{kuntschner11}. The prism designs use high transmission BaF2 as substrate, employed at an incidence angle of 45 degrees, with a relatively flat transmission of 0.92$\pm$0.01 over the wavelength ranges of interest. Prisms disperse with high transmission over a wide wavelength span, compared to gratings used in low order. These two spectrographs are identical except one extends to the edge of the K-band at 2.5~$\mu$m (Model A), which can sample the absorption bands of H$_{2}$O, CH$_{4}$, and CO and the continuum. The other (Model B) extends its wavelength range to the edge of the M-band at 5~$\mu$m, requiring it to be cooled to lower temperatures (85~K vs. 155~K for Model A) to reduce thermal background noise, to measure additional strong bands of H$_{2}$O, CO, and
CH$_4$, as well as CO$_{2}$. Model B's wavelength coverage also approximates JWST/NIRSpec's HgCdTe detector cutoff \citep{rauscher14}.

Both spectrographs have visible wavelength coverage down to 0.5~$\mu$m. Short-wavelength measurements probe an exoatmosphere's Rayleigh-scattering slope to measure its optically-thick radius \citep{tinetti10,Benneke12,Benneke13,griffith14,betremieux16,betremieux14,betremieux15,zellem15}.   These visible wavelengths also crucially probe regions where aerosols (clouds and hazes) absorb and scatter \citep[e.g.,][]{deming13,kreidberg14b,morello15,sing16}; the slope of the transmission spectrum across the visible can constrain the scattering particle size. In addition to the optical wavelength's importance in determining particulate scattering hazes, it is also important for constraining the abundances of strong optical/UV absorbers such as the alkali metals and metal oxides/hydrides (Na, K, TiO, VO, and FeH for
the hottest planets). Recent studies suggest the presence of clouds is common in exoplanet atmospheres \citep{sing16,iyer16,stevensonhst16}, which can partially conceal the spectral modulation of chemical absorbers within the atmosphere. If not properly accounted for, the decreased modulation can result in an underestimation of the atmosphere's molecular abundances \citep{Madhusudhan14,benneke15,iyer16}.

Both spectrographs feature continuous spectral coverage across their entire bandpass. This instantaneous wavelength coverage mitigates measurement uncertainty when combining observations conducted with different instrument modes or at different times. For example, two instruments can have different bias offsets which, if improperly treated during reduction, could alter the measured transit depths and derived quantities \citep[e.g.,][]{line16}. Alternatively, activity of the host star can change the measured transit depth \citep{pont08,agol10,carter11,desert11,sing11,ballerini12,oshagh14,fraine14,kreidberg14a, kreidberg14b,kreidberg15,damasso15,zellem15,sing16}, although the effect is negligible in most cases \citep{zellem17}.

The resolutions of both spectrographs vary across the wavelength band with a
minimum value of $R>34$ at 1.4~$\micron$. This resolution is sufficient
to resolve the H$_{2}$O bands, detect 
continuum, and to remove potential confusion from other
molecular species such as CO, CO$_{2}$, and CH$_{4}$. As indicated by the simulations we present here, this spectral resolution is sufficient to make precise abundance measurements of CO, CO$_2$, CH$_4$, and
H$_{2}$O (see Section~\ref{sec:discuss}).

\subsection{Telescope}
We adopt a telescope primary mirror diameter of 0.4~m with negligible attenuation; this mirror size was chosen to reflect the design of the successful WISE telescope \citep{liu08, wright10}. The use of a high-throughput prism
spectrograph and the availability of bright transiting exoplanet
targets, both currently-known and those predicted to be discovered with NASA's Transiting Exoplanet Survey Satellite \citep[TESS;][]{sullivan15}, make this telescope size viable for atmospheric
characterization.


The telescope has an orbit that allows for continuous measurement of each transit lightcurve without any interruptions due to, for example, Earth occultation. This orbit minimizes
measurement systematics associated with reacquisition of the target \citep[e.g. the \textit{Spitzer} and \textit{Hubble} ramp and hook;][]{beerer11, sing11, todorov12, knutson12, lewis13, kreidberg14a, kreidberg14b, zellem14, kreidberg15}.



\subsection{Exoplanetary Systems}
For an initial input spectrum we use a K5 star with a 9.4~H-mag. This stellar type is approximately  representative of the average stellar type that would be discovered by the TESS transiting exoplanet survey mission \citep{sullivan15}. The star is modeled as a black body with a stellar temperature of 4480 K. Using this host star type, we then step through a grid of primary star apparent magnitudes from 9.4 to 1.4~H-mag; this range of host star magnitudes was chosen to sample a majority of the currently-known exoplanets and those projected to be discovered by TESS \citep{sullivan15, zellem17}.


We simulate five different exoplanet archetypes defined by their mean molecular weights $\mu$ and terminator temperatures $T$ (Table~\ref{tbl-1}): a H/He-dominated hot Jupiter ($\mu = 2.3$~amu, T~$=1500$~K), a H/He-dominated warm Neptune ($\mu = 2.3$~amu, T~$=700$~K), a N$_{2}$-dominated warm Neptune ($\mu = 28$~amu, T~$=700$~K), a H/He-dominated sub-Neptune ($\mu = 2.3$~amu, T~$=600$~K), and a N$_{2}$-dominated sub-Neptune ($\mu = 28$~amu, T~$=600$~K). While a N$_2$-dominated exoplanet atmosphere is unlikely, we use it here as a proxy within these simulations for a spectroscopically-inactive high mean molecular weight gas, thereby leading to a much smaller scale height. Given that the bulk atmospheric compositions for ``super-earth'' type exoplanets can range anywhere from H$_2$ to CO$_2$ dominated ($\mu = 2$~amu to $\mu = 44$~amu, respectively), we choose a conservative value that lies somewhere in-between.

  
\subsection{Radiometric model}
\begin{figure*}
\centering
\includegraphics[width=0.75\textwidth]{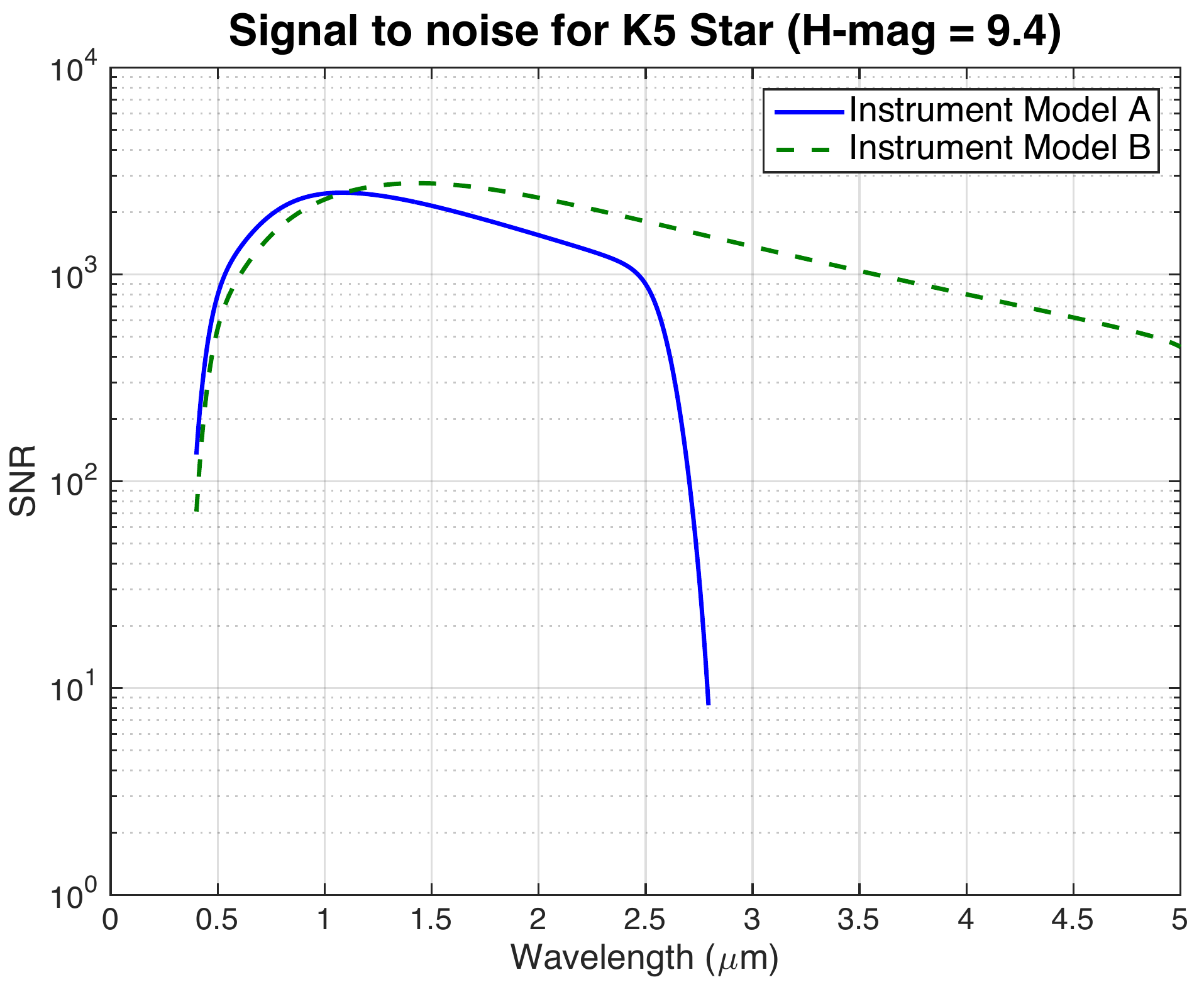}
\caption{\label{fig:radmodel}The SNR predicted for the 9.4~H-mag K5 host star over the duration of the transit with 
  the radiometric model for Instrument Model A (0.4 -- 2.5 $\mu$m; blue solid line) and Instrument Model B (0.4--5 $\mu$m; green dashed line).
  The parameters used in the radiometric model for each Model are given in
  Table~\ref{tab:rm}.}
\end{figure*}

\begin{figure*}
\centering
\begin{minipage}{.5\textwidth}
  \centering
  \includegraphics[width=1.\linewidth]{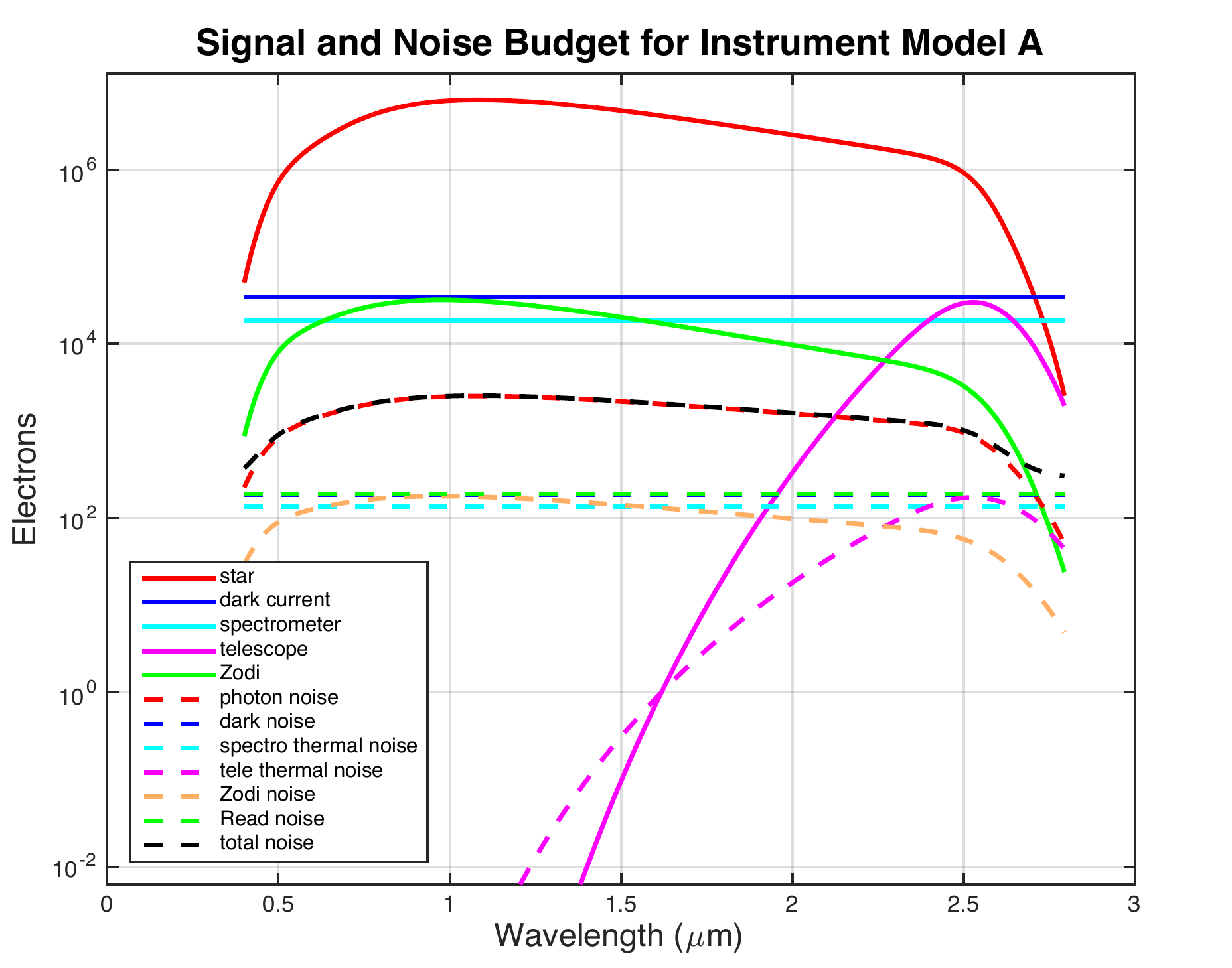}
  \label{fig:test1}
\end{minipage}%
\begin{minipage}{.5\textwidth}
  \centering
  \includegraphics[width=1.\linewidth]{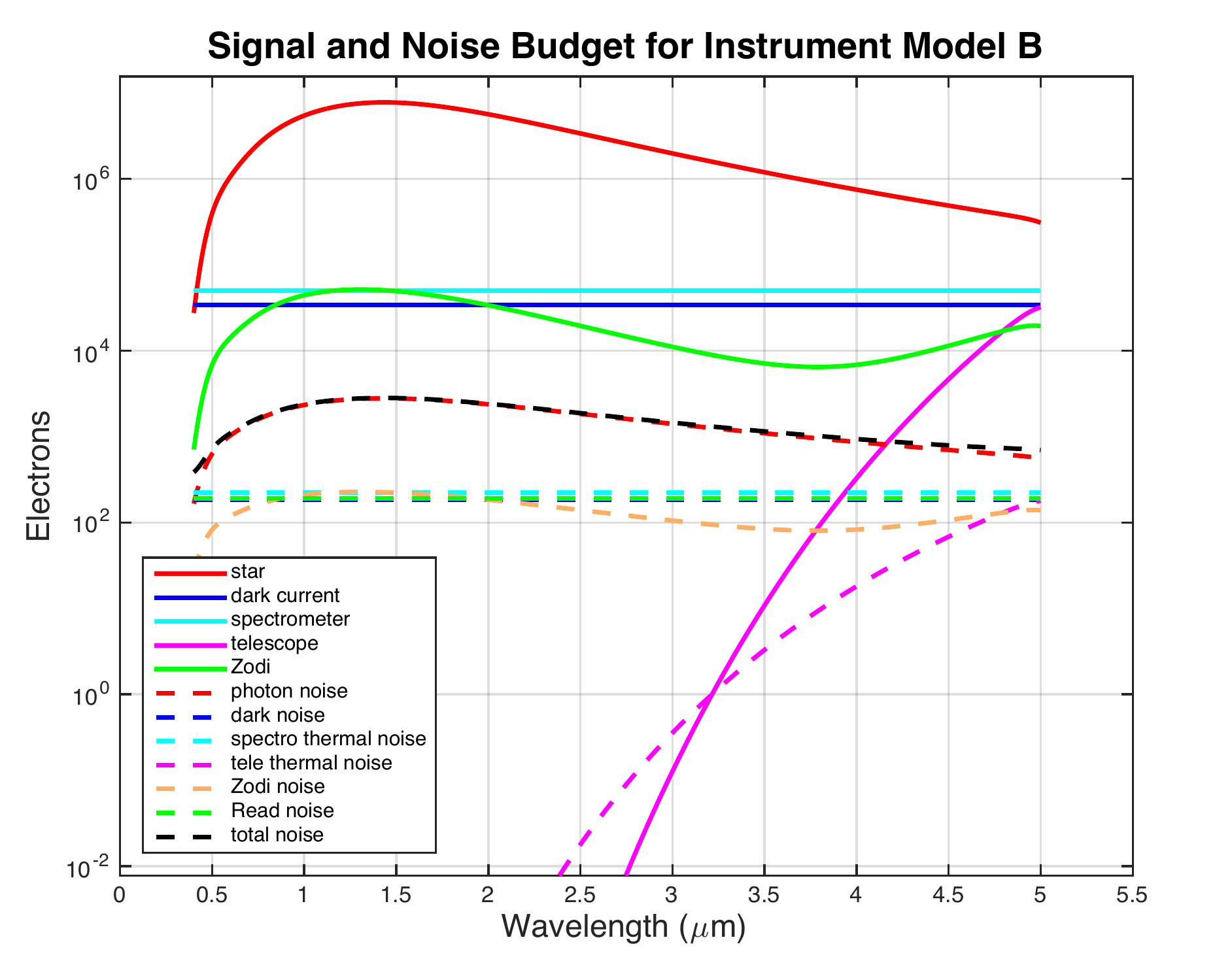}
  \label{fig:test2}
\end{minipage}
\caption{The signal and noise budget for the 9.4~H-mag K5 host star over the duration of the transit with the radiometric model for Instrument model A (left) and Instrument model B (right). The parameters used in the radiometric model for each case are given in
  Table~\ref{tab:rm}.}
  \label{fig:budget}
\end{figure*}

Signal-to-noise ratios (SNR; Fig.~\ref{fig:radmodel}) are
assessed using a radiometric model (Table~\ref{tab:rm}) assuming a one hour transit duration. The signal from the transiting exoplanet is followed through the spectrometer and the
expected signal and noise per pixel is calculated.
Included noise sources are the shot noise, the zodiacal background, background noise from the telescope, background noise from the instrument, detector dark current, and read-out noise (Fig.~\ref{fig:budget}). For both spectrographs, the noise is
dominated by the photon noise.  These SNRs were then used in the generation of synthetic transit data (Section~\ref{sec:retrieval}).

%
%
%
%
%
%
%

\begin{figure*}
\centering
\includegraphics[width=0.75\textwidth]{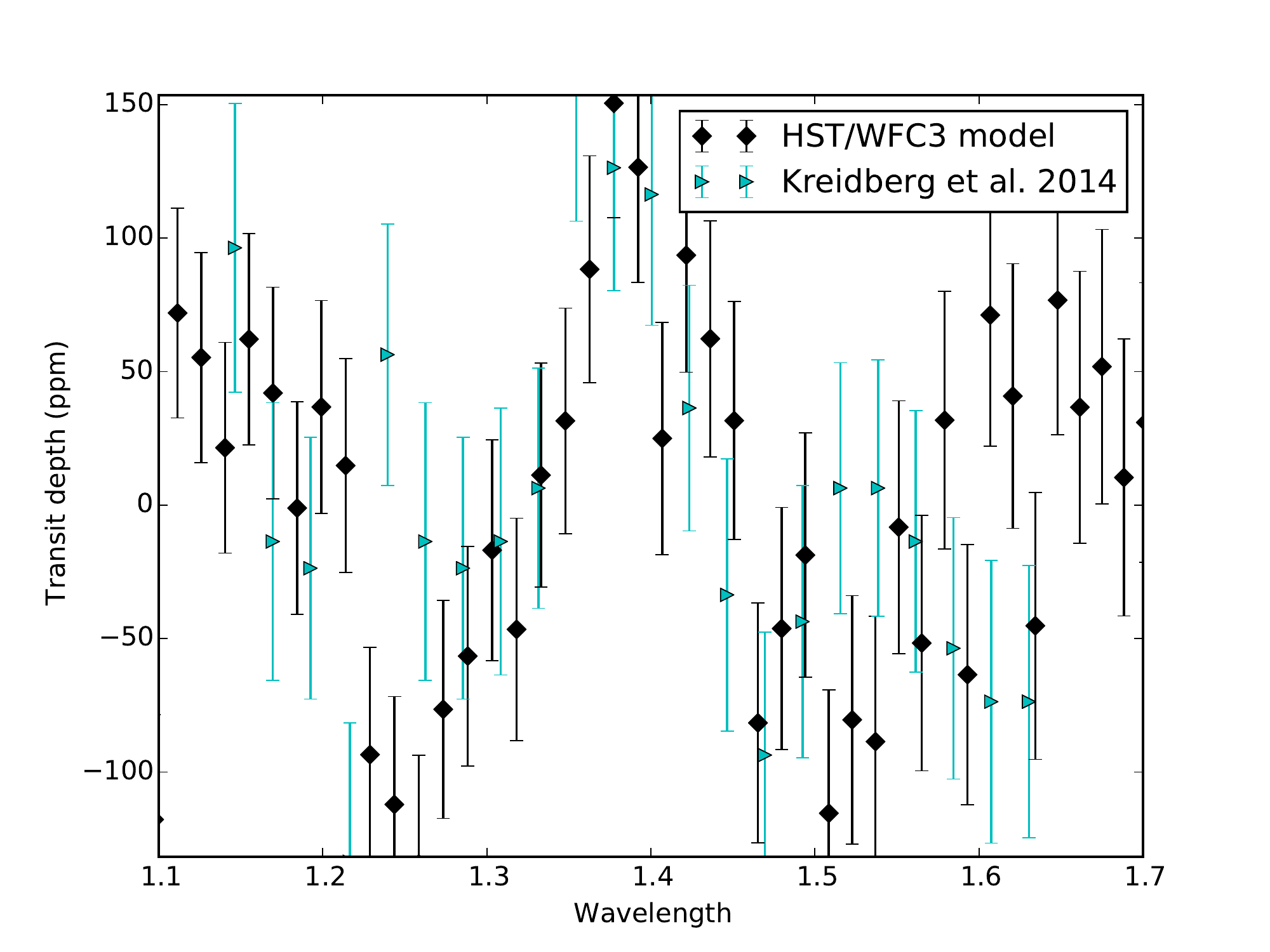}
\caption{\label{fig:model_vs_kreidberg}Our radiometric instrument model is validated by comparing our simulated HST/WFC3 observations of WASP-43b (black diamonds) and its published transmission spectrum (teal triangles).}
\end{figure*}

With our radiometric model, we also produced a simulation of WASP-43b as observed with Hubble/WFC3 to compare the model's performance with real observations \citep{kreidberg14b}. Our radiometric model is validated by excellent agreement with WASP-43b's measured transmission spectrum (Figure \ref{fig:model_vs_kreidberg}).

\section{\label{sec:retrieval}Retrieval of atmospheric abundances}
To determine the information content of an exoplanet atmosphere as a function of its apparent H-magnitude, we first generated forward models of the exoplanet's transmission spectrum assuming solar metallicity, equilibrium chemistry, and an isothermal temperature-pressure profile at the terminator with the CHIMERA radiative transfer free-retrieval code \citep{line13,line13b,swain14,line14,kreidberg15,morley16}. To simulate a real observation, realistic uncertainties simulated from the radiometric model were added to this forward model by drawing from a Gaussian distribution, assuming neighboring resolution elements have uncorrelated errors \citep[e.g.,][]{kreidberg14b}.

The transmission spectra are generated with 12 free parameters, six of which are the volume mixing ratios of the molecular gases H$_2$O, CH$_4$, CO, CO$_2$, NH$_3$, NaK, and N$_2$. The remaining five parameters are as follows: first, an effective temperature (T), which directly impacts the amplitude of the spectral features. Second, we use a scaling to the fiducial 10 bar planet radius ($x$R$_p$). This parameter manifests itself as a zero-point offset in the spectra as well as a minor impact on the amplitude of the features. Third, we include an opaque gray cloud parameterized with a cloud-top pressure (P$_c$). We set the cloud-top pressure to $\sim$31 bars, effectively simulating a  cloud-free atmosphere. The atmospheric transmittance at atmospheric levels deeper than the cloud top is set to zero. Clouds as observed with WFC3 tend to mute features by a factor of $\sim$1/2 \citep{iyer16}. This certainly would have an impact on the feature SNR and our ability to determine precision abundances. However, it is also possible that clouds need not be gray at all wavelengths. Some cloud opacities, if small enough particles, tend to wane quickly beyond a few microns \citep{howe12, morley12}.  We also model hazes in an ad-hoc fashion using the following approximation \citep{etangs08}:
	\begin{equation}
	\sigma = \sigma_0\left(\frac{\lambda}{\lambda_0}\right)^{-\beta}
\end{equation} 
where $\sigma_0$ is the magnitude of the haze cross-section relative to H$_2$O Rayleigh scattering at 0.4 $\mu$m, and $\beta$ is the wavelength power law index that describes the slope.

\begin{figure*}
\centering
\epsscale{1.0}
\includegraphics[width=1\textwidth]{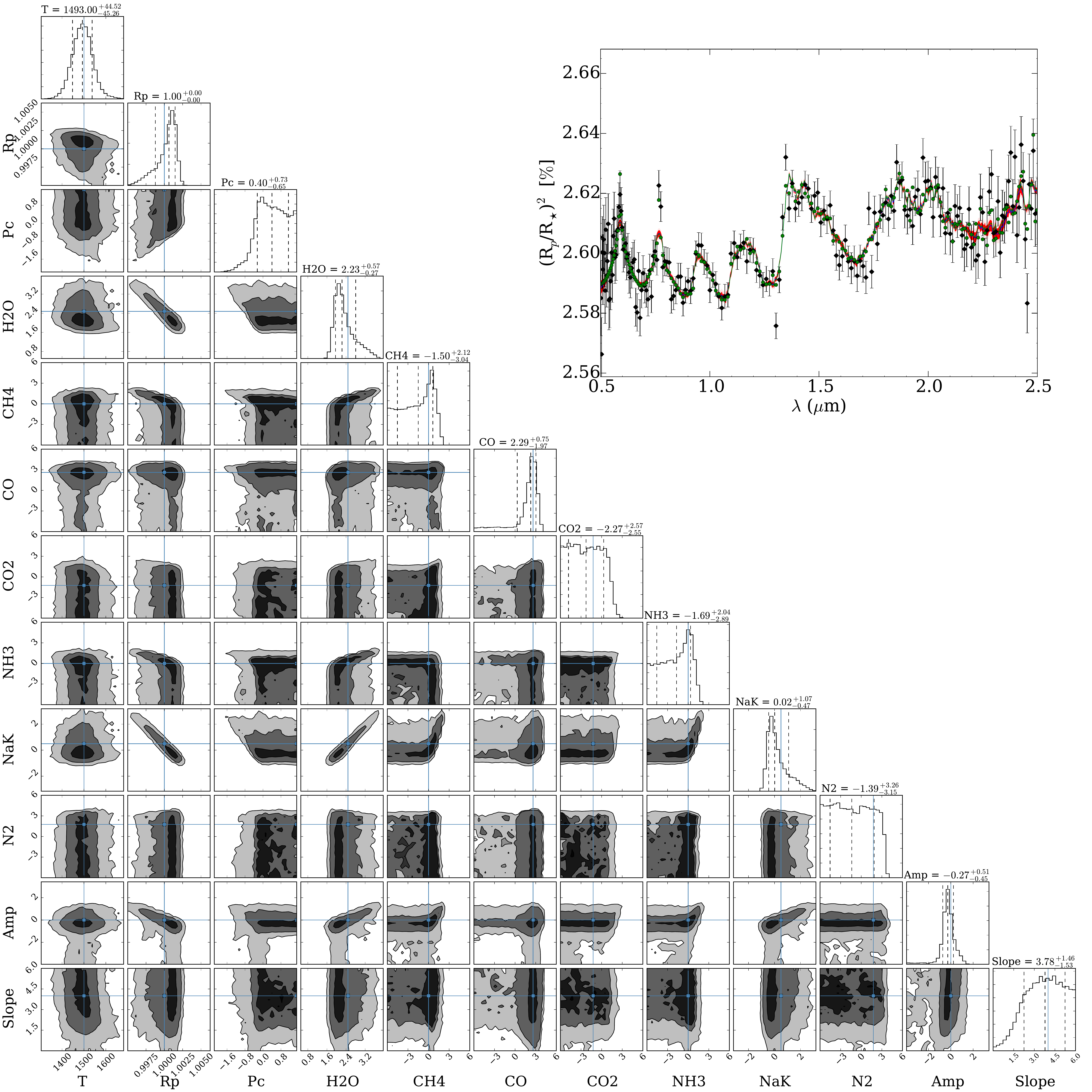}
\caption{\label{fig:triangleA}Posterior distributions of model
  parameters for a low mean molecular weight hot Jupiter assuming an ostensibly cloud free atmosphere, wherein clouds set to $\sim$31 bars within the forward model. The 1, 2, and 3$\sigma$ surfaces are shown shaded in black, dark grey, and light grey, respectively. The example shown is for a 5.4~H-mag host star and a spectrograph covering $0.4$ to $2.5~\micron$ (Model A). For the associated spectrum (inset), the black diamonds are the simulated data points and the green circles are the retrieved data points. The best fit spectrum is shown in green, while the blue line depicts the median of the 1000 best fit spectra generated by the Markov chain Monte Carlo approach. The $\pm$1$\sigma$ and $\pm$2$\sigma$ spread in the spectra are shown
in dark- and light-red, respectively.} 
\end{figure*}

\begin{figure*}
\centering
\epsscale{1.0}
\includegraphics[width=1\textwidth]{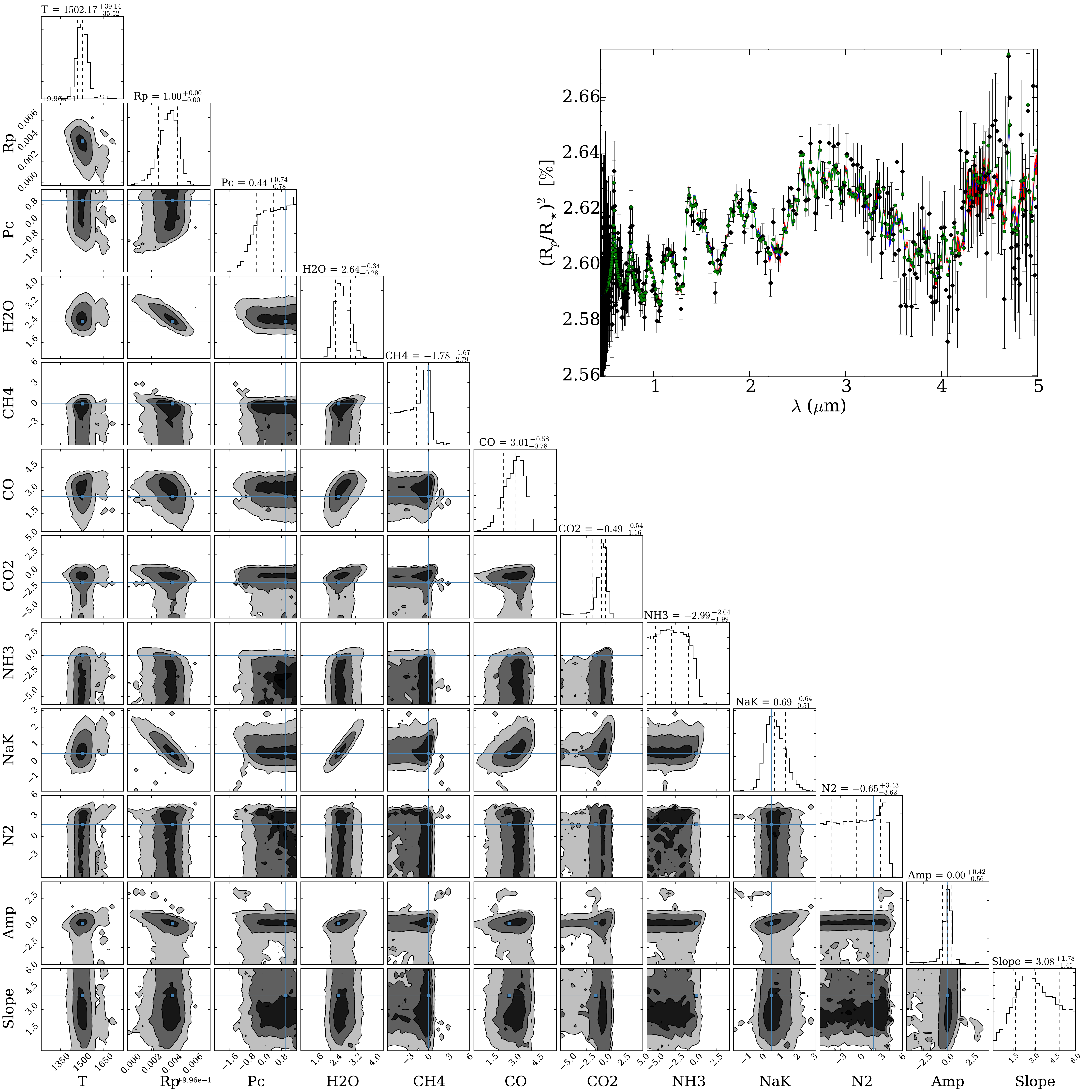}
\caption{\label{fig:triangleB}Same as Figure~\ref{fig:triangleA}, except using a spectrograph covering $0.4$ to $5.0~\micron$ (Model B).} 
\end{figure*}

We then retrieved each planet's molecular abundances with a full
Markov Chain Monte Carlo \citep[MCMC,][]{ford04} spectral retrieval on a total of 12 free model parameters (Table \ref{tab:params}), generating marginalized posterior distributions
for each of the atmospheric model parameters (Figures~\ref{fig:triangleA} and \ref{fig:triangleB}).  The widths of these posteriors represent the uncertainties of each retrieved parameter and are thus adopted as their degree of constraint. The results of this analysis are summarized in
Tables~\ref{tab:results} and \ref{tab:results_HMMW} where we compare the degrees of molecular abundance constraint of Models A and B as a function of host star brightness and planet type.

\begin{deluxetable}{c|c|c|ccccccc|c|c}
\rotate
\tablewidth{0pt}
\tabletypesize{\scriptsize}
\tablecaption{Retrieval free parameters and associated log-flat prior ranges. \label{tab:params}}
\tablehead{Equilibrium  & Radius  	&	Cloud-top  & \multicolumn{7}{c}{Molecular}		&	Rayleigh Scattering  &  Rayleigh Scattering \\
 Temperature & Scaling 	&	Pressure & \multicolumn{7}{c}{Abundance} &  Amplitude &   Slope\\
\tableline
\\
T(K)  & $x$R$_p$ & $\log$P$_c$(bar) & $\log$H$_2$O  & $\log$CH$_4$ & $\log$CO & $\log$CO$_2$ & $\log$NH$_3$ & $\log$NaK & $\log$N$_2$ &  Amp & Slope}
\startdata
0--3000  &  0.5--1.5	&	-4--1.5 &		-6--6&		-6--6	& -6--6	&		-6--6	&
-6--6 &		-6--6	& -6--6 & -5--5 & 0--6\\
\enddata
\end{deluxetable}

\clearpage
\begin{deluxetable}{c||cccc|cccc|cccc}
\rotate
\tabletypesize{\scriptsize}
\tablewidth{0pt}
\tablecolumns{21}
\tablecaption{Ratio of Uncertainties of Instrument Model A (0.5 to 2.5~$\mu$\textrm{\normalfont m}) to Model B (0.5 to 5~$\mu$\textrm{\normalfont m}) as a Function of Host Star Apparent Magnitude, Exoplanet Type, and Exoplanet Molecule.\label{tab:results}}
\tablehead{\colhead{} & \multicolumn{4}{c}{Hot Jupiter} & \multicolumn{4}{c}{Warm Neptune} & \multicolumn{4}{c}{Sub-Neptune} \\
\colhead{} & \multicolumn{4}{c}{T=1500 K, $\mu$=2.3} & \multicolumn{4}{c}{T=700 K, $\mu$=2.3} & \multicolumn{4}{c}{T=600 K, $\mu$=2.3} }

\startdata
H-mag & 
H$_2$O  & CH$_4$ & CO & CO$_2$& 
H$_2$O  & CH$_4$ & CO & CO$_2$& 
H$_2$O  & CH$_4$ & CO & CO$_2$\\
\tableline  
%

%
%
%
%
%
%
%
%

9.4 	&	3.16E+05 &	2.75E+01 &	5.75E+00 &	9.77E-01
	&	1.10E+05 &	6.61E+00 &	1.41E+01 &	2.57E+02
	&	1.91E+00 &	1.70E+00 &	2.51E+01 &	4.57E+02

\\

8.4 	&	5.89E-01 &	2.45E-02 &	1.12E-01 &	2.95E-01
	&	2.51E+01 &	3.31E-01 &	6.03E-01 &	1.02E+01
	&	1.05E+00 &	1.15E+00 &	1.07E+03 &	2.69E+01
\\

7.4 	&	1.05E+00 &	2.45E+01 &	2.57E+01 &	7.24E+01 &
		9.12E-01 &	1.78E+00 &	4.57E+00 &	5.75E+02 &
		
		1.05E+00 &	1.15E+00 &	1.07E+03 &	2.69E+01 \\
6.4 	&	4.17E-01 &	7.94E+00 &	1.29E+01 &	6.61E+00 &
		5.50E+00 &	4.37E+00 &	3.89E+02 &	1.20E+01 &
		
		1.20E+00 &	1.26E+00 &	1.91E+03 &	1.07E+02 \\
5.4 	&	3.98E+00 &	5.13E+00 &	1.55E+04 &	6.76E+01 &
		6.17E+00 &	3.24E+00 &	6.31E+01 &	1.23E+01 &
		
		1.12E+00 &	1.10E+00 &	2.82E+03 &	6.46E+01 \\
4.4 	&	1.70E+00 &	3.80E+00 &	2.75E+01 &	3.24E+03 &
		1.95E+00 &	1.95E+00 &	1.41E+03 &	5.50E+01 &
		
		1.15E+00 &	1.15E+00 &	1.45E+03 &	1.66E+02 \\
3.4 	&	1.91E+00 &	2.51E+01 &	5.25E+01 &	3.55E+01 &
		1.41E+00 &	1.32E+00 &	1.20E+03 &	7.94E+01 &

		1.15E+00 &	1.12E+00 &	3.09E+03 &	5.37E+02 \\
2.4 	&	1.26E+00 &	2.29E+04 &	1.70E+00 &	1.45E+01 &
		1.58E+00 &	1.48E+00 &	1.15E+03 &	2.95E+01 &

		1.05E+00 &	1.05E+00 &	2.95E+03 &	4.79E+01 \\
1.4 	&	9.33E-01 &	1.70E+04 &	1.17E+00 &	2.75E+04 &
		1.41E+00 &	1.35E+00 &	2.45E+04 &	5.01E+01 &

		1.05E+00 &	1.00E+00 &	5.50E+02 &	2.40E+02 \\
\hline
Median&	1.26E+00 &	2.45E+01 &	1.29E+01 &	3.55E+01
	&	1.95E+00 &	1.78E+00 &	3.89E+02 &	5.01E+01
	&	1.15E+00 &	1.15E+00 &	1.48E+03 &	1.66E+02
\\

\enddata
\end{deluxetable}

\clearpage
\begin{deluxetable}{c||cccc|cccc}
\rotate
\tabletypesize{\scriptsize}
\tablewidth{0pt}
\tablecolumns{21}
\tablecaption{Ratio of Uncertainties of Instrument Model A (0.5 to 2.5~$\mu$\textrm{\normalfont m}) to Model B (0.5 to 5~$\mu$\textrm{\normalfont m}) as a Function of Host Star Apparent Magnitude, Exoplanet Type, and Exoplanet Molecule.\label{tab:results_HMMW}}
\tablehead{\colhead{} & \multicolumn{4}{c}{Warm Neptune} & \multicolumn{4}{c}{Sub-Neptune} \\
\colhead{} & \multicolumn{4}{c}{T=700 K, $\mu$=28} & \multicolumn{4}{c}{T=600 K, $\mu$=28} }

\startdata
H-mag & 
H$_2$O  & CH$_4$ & CO & CO$_2$& 
H$_2$O  & CH$_4$ & CO & CO$_2$\\
\tableline  
%
9.4 &	1.95E+03 &	1.00E-01 &	7.41E+00 &	6.31E+00
	&	1.55E-01 &	3.24E+05 &	8.71E-02 &	5.62E+00	
\\
8.4 &	6.92E-01 &	4.47E+06 &	6.03E+00 &	2.51E+01
	&	2.24E+00 &	1.62E-04 &	9.55E-02 &	5.13E+05
\\
7.4 	&	2.24E+00 &	5.01E+02 &	4.79E-01 &	5.13E-01 &
 		2.95E-02 &	6.31E-01 &	1.82E+00 &	7.94E+04 \\
6.4 	&	4.07E-06 &	5.75E+05 &	1.12E+00 &	9.12E+04 &
		7.59E+00 &	2.57E+00 &	1.20E+00 &	1.02E+03 \\
5.4 	&	1.38E+00 &	3.24E+02 &	9.77E-01 &	4.79E-01 &
		3.47E+00 &	2.24E+00 &	5.37E-01 &	6.17E+00 \\
4.4 	&	2.82E-01 &	5.75E-01 &	1.00E+00 &	1.51E+03 &
		3.55E+00 &	2.51E+00 &	3.72E+01 &	9.12E+01 \\
3.4 	&	9.77E-01 &	9.77E-01 &	6.03E+00 &	2.57E+00 &
		2.40E+00 &	1.91E+00 &	3.39E-01 &	2.82E+00 \\
2.4 	&	1.78E+00 &	2.69E+00 &	2.88E-01 &	8.71E-01 &
		2.19E+00 &	1.58E+00 &	3.02E+01 &	2.19E+00 \\
1.4 	&	2.82E+00 &	2.51E+00 &	1.38E+00 &	3.55E+00 &
		2.57E+00 &	2.04E+00 &	9.12E+00 &	2.51E+00 \\
\hline
Median	&	1.38E+00 &	2.69E+00 &	1.12E+00 &	3.55E+00
		&	2.40E+00 &	2.04E+00 &	1.20E+00 &	6.17E+00 \\

\enddata
\end{deluxetable}

\begin{figure*}
\centering
\includegraphics[width=0.75\textwidth]{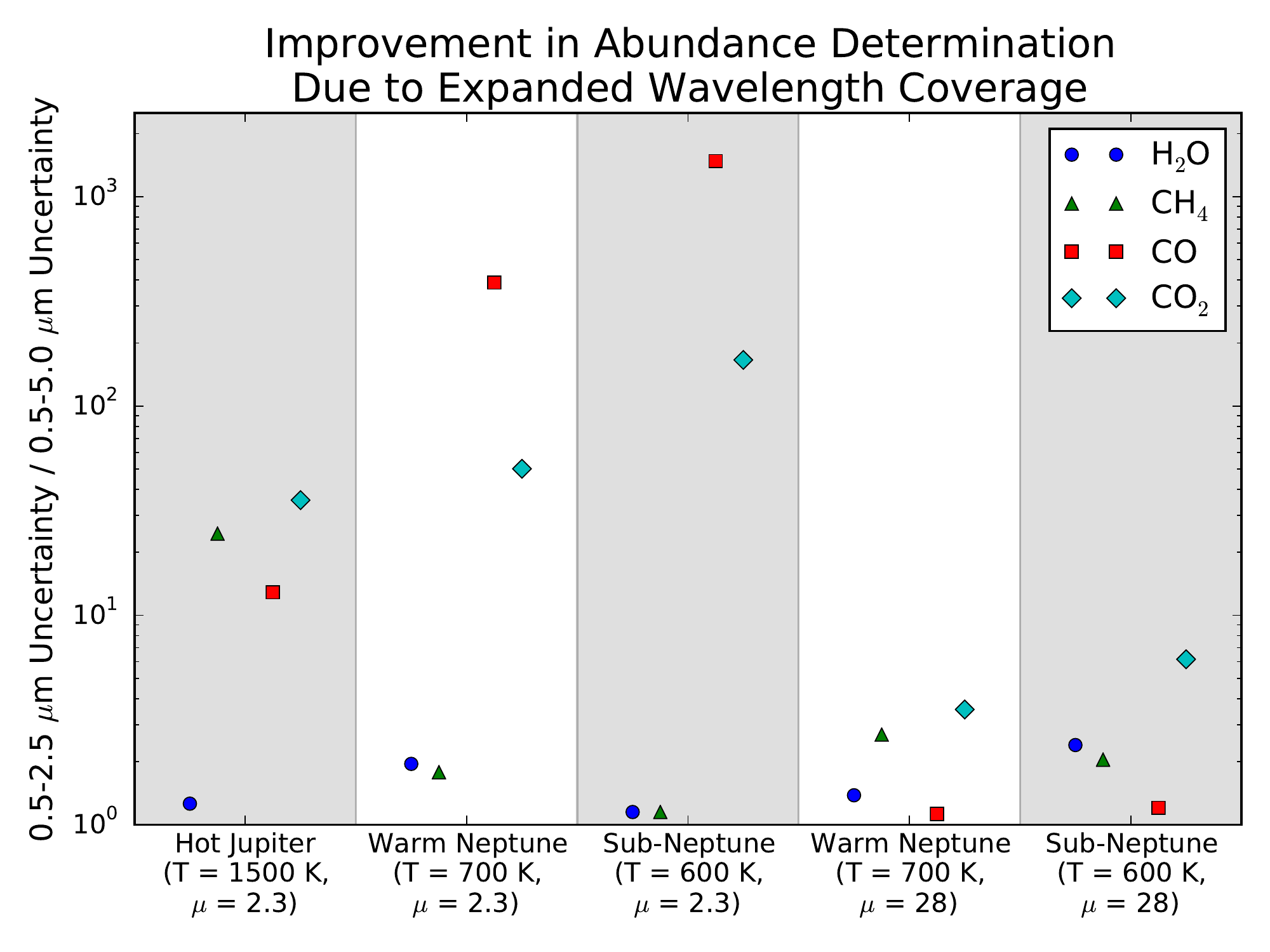}
\caption{\label{fig:final_fig_log} Ratio of measured uncertainties in the molecular abundances by Model A ($0.5$--$2.5\mu$m) vs. Model B ($0.5$--$5.0\mu$m) for each planet type in our study. The uncertainties shown are the median across all the magnitudes (to debias the estimator) and are plotted for each species and planet type. $10^n$ signifies an $n$ order of magnitude decrease in uncertainty due to additional wavelength coverage.}
\end{figure*}
\section{\label{sec:discuss}Discussion and Conclusions}
%
%



Our study indicates that spectral measurements extending from 0.5 to 5~$\micron$ wavelengths significantly improves the characterization of the major C and O molecules in exoplanetary atmospheres, in comparison to 0.5 to 2.5~$\micron$ (Tables~\ref{tab:results} and \ref{tab:results_HMMW}; Fig.~\ref{fig:final_fig_log}). Overall, due to the spectral sampling of CH$_4$ at 3.6~$\micron$, CO at 4.8~$\micron$, and CO$_2$ at 4.3~$\micron$, the measurement uncertainties of a low mean molecular weight atmosphere decrease up to $\sim$3 orders of magnitude for CO, $\sim$2 orders of magnitude for CO$_{2}$, and $\sim$1 order of magnitude for CH$_{4}$; for a high mean molecular weight atmosphere, the measurement uncertainties decrease up to a factor of $\sim$6 for CO$_{2}$ and a factor of $\sim$2 for CH$_{4}$, while the CO abundance uncertainties are largely unchanged. For both atmosphere types, the uncertainty in the H$_{2}$O abundance is largely unaffected due to its strong absorption band at $\sim$1.3--1.6~$\micron$. Thus an accurate carbon budget, used to calculate the C/O ratio for example, requires continuous wavelength coverage out to 5~$\micron$ \citep{greene16}. We find the increased wavelength coverage of 2.5 to 5.0 $\mu$m has a clear advantage for the determination of both carbon species and water in all planet cases simulated and typically improves molecular abundance uncertainties by 1 to 2 orders of magnitude (Tables~\ref{tab:results} and \ref{tab:results_HMMW}; Fig.~\ref{fig:final_fig_log}). These results indicate that JWST/NIRSpec, which has the capability to simultaneously cover 0.6--5~$\mu$m \citep{rauscher14}, or other future proposed missions consistent with the recent call for a dedicated exoplanet atmopshere survey telescope \citep{cowan15} such as ARIEL \citep{puig16,tinetti16} and FINESSE \citep{bean17}, would serve as excellent tools for investigating exoplanet atmospheres.

In addition, we find that increasing the brightness of the host star decreases molecular abundance measurement uncertainties due to the increased signal-to-noise ratio of the observations. Any deviation from this trend within our results is largely due to differing noise instantiations between simulated observations. If averaged over a large enough sample or successive transit observations, these deviations would decrease significantly.

\section{\label{sec:future}Future Work}
For future work, we plan on utilizing the chemically consistent version of the CHIMERA radiative transfer model \citep{line16} in place of the free retrieval model used in this study. That approach could prove more beneficial in that the chemically consistent model only retrieves total C/O ratio and metallicity, thereby resolving possible degeneracies existing within parameter-space and resulting in greater constraints on abundances. Although proper consideration of C/O ratios is best done in the
context of chemically consistent modeling due to the strong prior effect as described in \citet{line13a}, the additional wavelength coverage (from 2.5 to 5.0 $\mu$m) decreases the uncertainty in the C/O ratio by a factor of $\sim$1.7, except in the case of the high mean molecular weight atmosphere warm Neptune where no significant improvement is seen; trials with chemically consistent modeling indicate that the use of a non-chemically consistent code is limiting our ability to make reliable C/O estimates in this case in the presence of low feature SNR data. In a sense, atmospheric composition inference in those regimes will be largely prior driven.


\section*{Acknowledgments}
This work was conducted at the Jet Propulsion Laboratory, California Institute of Technology under contract with the National Aeronautics and Space Administration. $\textcopyright$ 2017. All rights reserved.

J.C. and R.T.Z. thank JPL's ExoSpec team for their helpful comments.


\bibliographystyle{apj}       
\bibliography{references}

\end{document}